\documentclass[usenatbib,usegraphicx,fleqn]{ctextemp_mn2e}
\usepackage{amsmath,amssymb,amsfonts,bm,color,multirow,multicol,array}
\usepackage{charter}   
\pdfoutput=1

\usepackage{graphicx}
\usepackage{natbib}
\usepackage{charter}
\usepackage[english]{babel}

\usepackage{ulem}

\newcommand{\mdot}{\ensuremath{\dot M}}
\newcommand{\ldot}{\ensuremath{\dot L}}

\newcommand{\msun}{\ensuremath{M_\odot}}

\newcommand{\comment}[1]{} 

\title[Comparisons of MHD Propeller Model]{Comparisons of MHD Propeller Model with Observations of Cataclysmic Variable AE Aqr}

\author[Blinova et al.]{\parbox{\textwidth}
{A. A. Blinova$^{1,2}$, M. M.~Romanova$^{1,2}$\thanks{E-mail of
corresponding author: \texttt{romanova@astro.cornell.edu}},  G.
V.~Ustyugova$^{3}$, A. V.~Koldoba$^{4}$,  R. V.
E.~Lovelace$^{1,2,5}$
}\vspace{0.4cm}\\
\parbox{\textwidth}{ 
$^{1}$Department of Astronomy, Cornell University, Ithaca, NY 14853-6801\\
$^{2}$Carl Sagan Institute, Cornell University, Ithaca, NY 14853-6801\\
$^{3}$Keldysh Institute for Applied Mathematics, Moscow, 125047,
Russia \\
$^{4}$Moscow Institute of Physics and Technology, Dolgoprudny, Moscow Region, 141700, Russia \\
$^{5}$Department of Applied and Engineering Physics, Cornell
University, Ithaca, NY 14853-6801}}

\begin{document}
\maketitle

\begin{abstract}

We have developed a numerical MHD model of the propeller candidate
star AE Aqr using axisymmetric magneto-hydrodynamic (MHD)
simulations. We suggest that AE Aqr is an intermediate polar-type
star, where the magnetic field is relatively weak and an accretion
disc may form around the white dwarf. The star is in the propeller
regime, and many of its observational properties are determined by
the disc-magnetosphere interaction. Comparisons of the
characteristics of the observed versus modelled AE Aqr star show
that the model can explain many observational properties of AE
Aqr. In a representative model, the magnetic field of the star is
$B\approx 3.3\times10^5$ G and the time-averaged accretion rate in
the disc is $5.5\times 10^{16}$ g/s. Most of this matter is
ejected into conically-shaped winds. The numerical model explains
the rapid spin-down of AE Aqr through the outflow of angular
momentum from the surface of the star to the wind, corona and
disc. The energy budget in the outflows, $9\times 10^{33}$erg/s,
is sufficient for explaining the observed flaring radiation in
different wavebands.
 The time scale of ejections into
the wind matches the short time scale variability in the light
curves of AE Aqr.

\end{abstract}

\section{Introduction}

 AE Aqr is  a nova-like cataclysmic variable (CV).
 It consists of a magnetic white
  dwarf and a late-type companion star with a spectral type of K3–-K5.
The binary has a relatively long period of 9.88 hours. It is
widely believed that the companion star's atmosphere fills its
Roche lobe and matter flows out
 of the companion's Roche lobe to the white dwarf \citep{CasaresEtAl1996}.
Pulsations with a period of 33.08 s were detected in optical
\citep{Patterson1979}, UV (e.g., \citealt{EracleousEtAl1994}),
soft X-ray \citep{PattersonEtAl1980}, and hard X-ray
\citep{KitaguchiEtAl2014} wavebands.

AE Aqr could be classified as a typical intermediate polar (DQ
Her) type CV, except that it has a number of unusual properties:
 (1) It shows
 flaring radiation
in optical, ultraviolet and X-ray bands, which are all correlated
with each other (e.g., \citealt{Patterson1979, MaucheEtAl2012});
(2) Flaring radiation in the radio band shows a non-thermal
spectrum typical for electrons radiating in a magnetized plasma
\citep{BastianEtAl1988}. Authors compared the radio flares in
AE Aqr with those of the micro-quasar Cyg X-3; (3) The star is
spinning down rapidly, at a rate of $\dot P=5.64\times10^{-14}$ s
s$^{-1}$, which corresponds to a high spin-down power of $\dot
E_{\rm sd}\approx 6\times 10^{33} I_{50}$ erg/s (where
$I_{50}=I/10^{50}{\rm gcm}^2$) \citep{deJagerEtAl1994}; (4) The
estimated accretion luminosity, $\dot E_{acc}\lesssim 10^{32}$
erg/s (e.g., \citealt{Mauche2006}), is
much lower than the spin-down power; (5) The $H_\alpha$ spectral
lines are strongly variable and indicate the presence of outflows.
The Doppler tomograms based on the analysis of these lines
\footnote{Note that Doppler tomography implicitly assumes that
emission is optically thin and bound to the orbital plane of the
binary (e.g., \citealt{SchwopeEtAl1999}). Therefore, Doppler
tomography is not the optimal tool for investigating the presence
of winds.}
are different from the tomograms of other intermediate polars,
which typically indicate the presence of an accretion disc (e.g.,
\citealt{MarshEtAl1990}). No signatures of a disc were observed in
the AE Aqr tomograms (see Fig. 10 from \citealt{WelshEtAl1998}).

Different models were proposed to explain the observational
properties of AE Aqr. In one type of models, it is suggested that
the magnetic field of AE Aqr is very large, $B_s\approx
5\times10^7$ G, and the white dwarf spins down due to the
magneto-dipole radiation, as in the case of pulsars (e.g.,
 \citealt{Ikhsanov1998,Ikhsanov2006}). In
the second type of models, it is suggested that the magnetic field
of the star is typical for that of intermediate polars, $B\lesssim
2\times10^6$ G \citep{Warner1995}, and the star spins down due to
the interaction of the rapidly-rotating magnetosphere with the
surrounding matter in the propeller regime (e.g.,
\citealt{EracleousHorne1996}).

The apparent lack of an accretion disc in the Doppler tomograms
led to the suggestion that an accretion disc does not form;
instead, matter flows from the secondary star in a stream of
blobs, which interact with the magnetosphere of the WD directly
without forming a disc (e.g.,
\citealt{King1993,WynnKing1995,WynnEtAl1997}).
 In their model, blobs of matter interact with the WD ballistically, and a
 diffusive term
 has been added to the equations to describe the
interaction of the blobs with the magnetosphere of the star. The
flaring radiation in different wavebands is explained by the
collisions of blobs during their exit path from the white dwarf
\citep{WelshEtAl1998}.

 \citet{MeintjesdeJager2000} suggested that a stream of matter
flowing from the secondary star interacts with the external layers
of the magnetosphere, loses its angular momentum and forms a ring
or a small disc around the white dwarf. The matter distribution in
this ring is strongly inhomogeneous. Recent 3D MHD simulations of
matter flow in AE Aqr have shown that a stream of matter flowing
from the secondary star collides with itself after the 1st turn
around the white dwarf, forming a ring around it that subsequently
forms a turbulent disc \citep{IsakovaEtAl2016}. Next, the matter
of the disc is ejected by the magnetosphere of the WD due to the
propeller mechanism. In their model, the magnetic field of the WD
is large, 50MG, which leads to a rapid ejection of the inner disc
matter by the large, rapidly-rotating magnetosphere.

Overall, it is reasonable to suggest that, in the cases of
 IP-scale magnetic fields, some kind of
disc may form around the white dwarf in AE Aqr. The disc may have
unusual properties compared with the discs around non-propelling
stars: it may be much smaller than standard discs in IPs, which
are thought to have sizes of  Roche lobes.  It may also be
temporary due to the propeller action.

The propeller regime is expected in AE Aqr, and various
observational properties may be connected with this regime.
However, the disc-magnetosphere interaction in the propeller
regime is a complex phenomenon, and has not been sufficiently
studied. It has been investigated in a few theoretical works on
spherical accretion (e.g., \citealt{IllarionovSunyaev1975}) and
disc accretion \citep{LovelaceEtAl1999}. However, only restricted
numerical models of the propeller regime have been developed so
far. \citet{WangRobertson1985} studied the propeller regime in
two-dimensional numerical simulations using polar coordinates, and
thus studied the processes in the equatorial plane. They observed
that the matter of the inner disc interacts with the magnetosphere
of the star due to the magnetic interchange instability (see also
\citealt{AronsLea1976}). No outflows were observed due to the
two-dimensional polar geometry of their coordinate system.

The propeller regime has also been studied in axisymmetric
simulations
\citep{UstyugovaEtAl2006,RomanovaEtAl2005,RomanovaEtAl2009,RomanovaEtAl2018,LiiEtAl2014}.
These simulations have shown that the disc-magnetosphere
interaction is a strongly non-stationary process, where the inner
disc oscillates and most of the matter is ejected into the
outflows from the disc-magnetosphere boundary, while a smaller
amount of matter accretes onto the star. The WD spins down due to
the outflow of angular momentum into the matter- and
magnetically-dominated winds. This model can potentially explain
the different observational properties of CV AE Aqr. In this
paper, we have developed a propeller model of a star with
parameters corresponding to those of AE Aqr, calculated the
properties of the modelled star and compared them with the
observed properties of AE Aqr.

In Sec. \ref{sec:model}, we describe our  model of AE Aqr. In Sec.
\ref{sec:comparisons-observations}, we compare our model with the
observations. In Sec. \ref{sec:conclusions}, we summarize our
results. In Appendix \ref{app:numerical-model}, we provide the
details of our numerical model.

\begin{figure*}
\centering
\includegraphics[width=16.0cm]{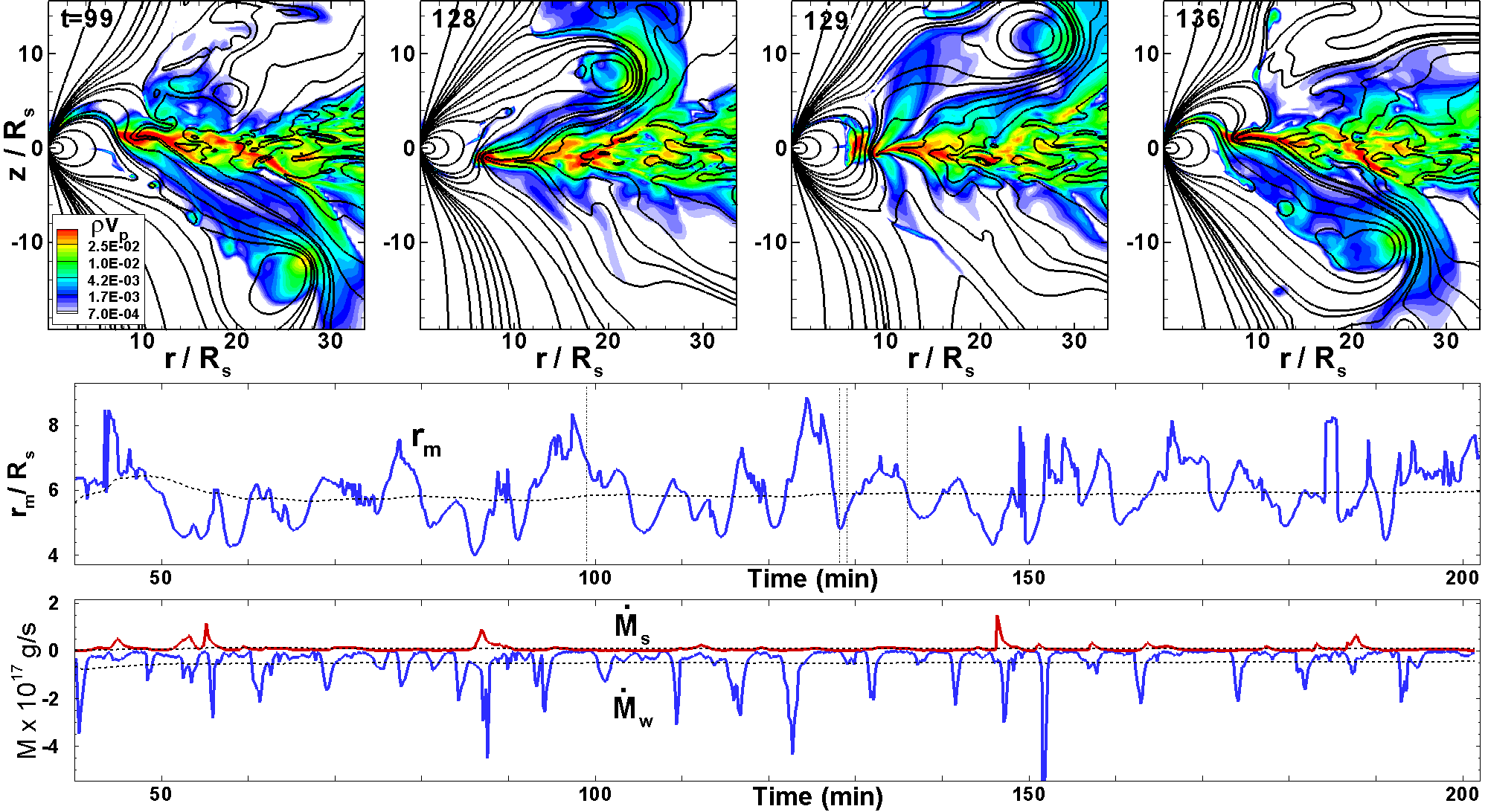}
 \caption{\textit{Top row:} A two-dimensional picture of matter flow in the model $\tilde{\mu}60$, shown at several moments in time.
 The colour background shows
  the matter flux density and the lines are sample
field lines. \textit{Middle row:} Variation of the inner disc
radius, $r_m$, scaled to the radiua of the star, $R_s$.
\textit{Bottom row:} Variation of matter fluxes at the stellar
surface, $\dot M_s$, and into the wind, $\dot M_w$. Dashed lines
show the time-averaged values. Vertical lines show the moments in
time corresponding to the top panels. } \label{2d-fluxes}
\end{figure*}

\section{Propeller model of  AE Aqr}
\label{sec:model}

\subsection{Preliminary estimates}

To model the propeller regime in AE Aqr, we take the mass and
radius of the white dwarf to be  $M_s=0.8 M_\odot$ and $
R_s=7\times10^3$ km, respectively (e.g.,
\citealt{MaucheEtAl2012}).
 The period of stellar rotation, $P_s=33.08$s, corresponds to
a corotation radius of
\begin{align}
r_{\rm cor} = \bigg(\frac{GM_s P_s^2}{4\pi^2}\bigg)^{1/3} \approx
1.44\times 10^4 \rm{km}\approx 2.05 R_s~. \label{eq:corotation
radius}
\end{align}

We suggest that a
disc forms around the WD. The luminosity
associated with accretion onto the stellar surface is low, and is
estimated to be  $\dot E_s\lesssim 10^{32}$ erg/s (e.g.,
\citealt{Mauche2006}). It is connected with the matter flux onto
the star as
\begin{equation}
\dot E_s = \eta \frac{\dot M_s GM_s} {R_s} \approx
1.53\times10^{31} \eta \dot M_{s15} \frac{\rm erg}{\rm s}~,
\label{eq:Edot_X from M_x}
\end{equation}
where  $\dot M_{s15}=\dot M_s/10^{15}$ g/s and the coefficient
$\eta$ takes into account the fact that matter may fall from the
finite distance $r$.  For example, if matter falls from the
distance of the corotation radius, $r_{\rm cor}\approx 2 R_s$,
then $\eta = 0.5$.  The corresponding accretion rate onto the star
is:

\begin{equation}
\dot M_s=\eta \frac{\dot E_s R_s}{GM_s} \approx 1.31 \times \eta
10^{15} \frac{\dot E_{s32}}{\eta_{0.5}} \frac{\rm g}{\rm s}~,
\label{eq:Mdot-from Lx}
\end{equation}
where $\dot E_{s32}=\dot E_s/10^{32}{\rm erg/s}$ and
$\eta_{0.5}=\eta/0.5$.

The accretion disc is stopped by the magnetosphere of the star at
the radius of $r_m$, where the matter stress in the disc is equal
to the magnetic stress in the magnetosphere. This condition can be
approximately described by the formula for the Alfv\'en radius,
obtained for non-rotating stars (e.g.,
\citealt{PringleRees1972,GhoshLamb1978}):
\begin{align}
r_m=k_m \frac{\mu_s^{4/7}}{(2GM_s{\dot M_d}^2)^{1/7}}
\nonumber \\
 \approx 2.52\times10^4 k_m B_5^{4/7}{\dot M}_{d17}^{-2/7}~\rm{km} ~,
 \label{eq:magnetospheric-radius}
\end{align}
where  $\mu_s$ is the magnetic moment of the WD and $\dot M_d$ is
the accretion rate in the disc (which, in the propeller regime, is
larger than the accretion rate onto the star, $M_s$); $k_m\sim 1$
is a dimensionless coefficient. $B_5=B_s/10^5 \rm{G}$ is the
normalized magnetic field of the WD, and ${\dot M}_{d17}=\dot
M_d/10^{17}\rm{g/s}$  is the normalized accretion rate in the
disc.

To estimate the magnetospheric radius in AE Aqr, one needs to know
the accretion rate $\dot M_d$, which is not well known
\footnote{For example, \citet{PearsonEtAl2003} estimated the mass
transfer rate from the donor star as $\dot M=3\times 10^{17}$
g/s.}. We suggest that $k_m=1$ and the accretion rate in the disc
is ${\dot M}_d = 10^{17}$ g/s to obtain the magnetospheric radius,
$r_m\approx 3.60 B_5^{4/7} R_s$. Taking two reference values for
the magnetic field of IPs, $B_s=10^5$ G and $B_s=10^6$ G, we
obtain $r_m\approx 3.60 R_s$ and $r_m\approx 13.43  R_s$,
respectively. In both cases, the magnetospheric radius is larger
than the corotation radius, $r_{\rm cor}=2 R_s$, and the star is
in the propeller regime. However, the strengths of the propellers
are different.

The strength of a propeller is often measured by the fastness
parameter, $\omega_s$ (e.g. \citealt{Ghosh2007}), which is the
ratio of the angular velocity of the star, $\Omega_s$, to the
angular velocity at the inner disc, $\Omega_d$:
$\omega_s=\Omega_s/\Omega_d$. In the case of a Keplerian disc,
$\Omega_d=\sqrt{GM_s/r_m^3}$, and the fastness parameter can be
re-written in the following form:
\begin{eqnarray}
\omega_s=\bigg(\frac{r_m}{r_{\rm cor}}\bigg)^{3/2}\approx 2.32
B_5^{6/7}{\dot M}_{d17}^{-3/7} ~ .
\end{eqnarray}
Using the reference values for the accretion rate, ${\dot
M}_{d17}=\dot M_d/10^{17} \rm{g/s}$,  and for the magnetic field,
$B_s=10^5$ G and $B_s=10^6$ G,  we obtain the values of the
fastness parameter,
 $\omega_s\approx 2.3$ and $\omega_s\approx 16.6$, respectively.  These estimates
are helpful in restricting the values of the fastness parameter in
the modelling of AE Aqr.

\begin{table*}
\centering
\begin{tabular}{l|l|lll|ll|llll}
 Model &   $\tilde\mu$ & $\langle \tilde r_m \rangle$ & $\langle {\tilde r_m} \rangle/{\tilde r_{\rm cor}}$ & $\langle\omega_s\rangle $ & $\langle\widetilde{\dot{M}}_s\rangle$ & $\langle\widetilde{\dot{M}}_w\rangle$  & $f_{\rm eff}$  \\
\hline
$\tilde\mu30$   &  30   & 5.0 & 2.5 & 3.9         & 0.17 & 0.76            & 0.82   \\
\hline
$\tilde\mu60$   &  60   & 5.9 & 3.0 & 5.1         & 0.22 & 1.24            & 0.85   \\
\hline
\end{tabular}
\caption{Dimensionless values obtained in two models of the strong
propeller regime, where the corotation radius $r_{\rm cor}=2 R_s$.
Parameter $\tilde\mu$ determines the final values of the
time-averaged magnetospheric radii, $\langle \tilde r_m \rangle$,
and the fastness parameter, $\omega_s$. Here,
$\langle\widetilde{\dot{M}}_s\rangle$  is the time-averaged matter
flux onto the star;  $\langle\widetilde{\dot{M}}_w\rangle$ is the
matter flux into the wind, calculated through the surface
$S(r=10,z=\pm10)$ at condition $v_{\rm min}>0.1 v_{\rm esc}$;
$f_{\rm eff}$ is the propeller efficiency.}
 \label{tab:dimensionless
values}
\end{table*}


\subsection{Axisymmetric  MHD Simulations of AE Aqr}
\label{subsec:numerical model}

 As a base, we use the axisymmetric model of the
propeller regime developed in our group (e.g.,
\citealt{LiiEtAl2014,RomanovaEtAl2018}). We start with the initial
conditions that are expected in disc-accreting propelling stars:
the simulation region consists of (a) a star with mass $M_s$,
radius $R_s$ and magnetic field $B_s$; (b) an accretion disc that
is cold and dense, and has an aspect ratio of $H/r\approx 0.15$
($H$ is the initial half-thickness of the disc); (c) a low-density
and high-temperature corona, which occupies the rest of the
simulation region (i.e., the space above and below the disk and
the star; see left panel of Fig. 1 from
\citealt{RomanovaEtAl2018}). Initially, the matter in the disc and
in the corona are in rotational equilibrium, in which the density
and pressure distributions are derived from the balance of the
gravitational, pressure gradient and centrifugal forces
\citep{RomanovaEtAl2002}.

The disc is turbulent. The turbulence is driven by the
magneto-rotational instability (MRI, e.g.,
\citealt{BalbusHawley1991,StoneEtAl1996,Armitage1998,Hawley2000}),
which is initiated by a weak poloidal magnetic field placed inside
the disc. In our model, the accretion rate corresponds to an
effective $\alpha-$parameter of $\alpha_{\rm mag}\approx
0.02-0.04$. A diffusivity term has been added to the code, with
the coefficient of diffusivity constructed in analogy with
$\alpha-$viscosity \citep{ShakuraSunyaev1973}: $\eta_d=\alpha_d
c_s H$, where $c_s$ is the sound speed and $\alpha_d$ is a
dimensionless parameter.

 In our models, we
suggest that the 3D instabilities should provide high diffusivity,
and we use $\alpha_d=1$ at radii $r<7R_s$, where the disc
typically interacts with the magnetosphere, and $\alpha_d=0$ in
the rest of the disc (to avoid suppressing the MRI-driven
turbulence). The star rotates with an angular velocity of
$\Omega_s$ such that the magnetosphere rotates more rapidly than
the inner disc, and the star is in the propeller regime.

 The models are calculated
in dimensionless variables.
The conversion procedure for dimensionalization and other details
of the model are provided in Appendix \ref{app:reference values}.

To model AE Aqr, we take the corotation radius $r_{\rm cor}=2 R_s$
(see Eq. \ref{eq:corotation radius}), or $\tilde r_{\rm cor}=2$ in
dimensionless units (see Appendix \ref{app:reference values}),
 and calculate all models for this
corotation radius. Since the magnetic field of AE Aqr is not
known, we model propellers
of different magnetospheric radii $r_m$. To do so, we vary the
parameter $\tilde\mu$, which we call the dimensionless magnetic
moment of the star \footnote{In our dimensionless model, the
variation of parameter $\tilde\mu$  can be interpreted as either
the variation of the stellar magnetic moment, $\mu_s$, or the
variation of the accretion rate in the disc, $\dot M_d$. A
variation of both values leads to the variation in $r_m$ as $\sim
(\mu_s^2/\dot M_d)^{1/7}$. }. We have developed numerical models
for $\tilde\mu=30$ and $60$ and obtained different magnetospheric
radii (and therefore different fastness values $\omega_s$).

We observed that, in both models, the inner disc strongly
oscillates and most of the inner disc matter is redirected into
the outflows. The top panels of Fig. \ref{2d-fluxes} show the
matter flux density and sample field lines in the model
$\tilde{\mu}60$. One can see that the magnetic field lines inflate
and open, and some of the matter is ejected into conically-shaped
outflows. Matter is ejected
at approximately $45-50$ degrees relative to the rotational axis.
Ejections above and below the equatorial plane alternate.
\footnote{In our axisymmetric simulations, matter is ejected into
an azimuthally-symmetric cone. In a more realistic,
three-dimensional world, we would expect non-axisymmetric
ejections of blobs of matter, which would shoot out in different
azimuthal directions.}

The disc-magnetosphere interaction occurs in a cyclic manner,
where episodes of matter accumulation in the inner disc are
followed by events of matter ejection into the winds and
simultaneous accretion of matter onto the star (see, e.g.,
\citealt{LiiEtAl2014}, see also \citealt{GoodsonEtAl1997}).

We find the dimensionless magnetospheric radius $\tilde r_m$ from
the simulations using the balance of magnetic and matter stresses,
$\tilde{B}^2/{8\pi}=\tilde{P}+\rho \tilde{v}^2$, where $\tilde B$,
$\tilde P$ and $\tilde\rho$ are the magnetic field, pressure and
density of matter in the equatorial plane, respectively
\citep{RomanovaEtAl2018}. This formula provides the instantaneous
value of $\tilde r_m$. The middle row of Fig. \ref{2d-fluxes}
shows that the inner disc strongly oscillates and the
magnetospheric radius $\tilde r_m$ varies with time, which is why
we calculate the time-averaged magnetospheric radius,
 \begin{equation}
\langle \tilde r_m(t)\rangle = \frac{\int_{t_i}^{t} dt' \tilde
r_m(t')}{\int_{t_i}^{t} dt'} ~ \label{eq:timeavg-rm}~.
\end{equation}
The dashed line in the middle row of Fig. \ref{2d-fluxes} shows
the time-averaged value $\langle\widetilde r_m\rangle$. We also
calculated the time-averaged fastness parameter,
\begin{equation}
\langle\omega_s\rangle = \bigg(\frac{\langle \tilde r_m
\rangle}{\tilde r_{\rm cor}}\bigg)^{\frac{3}{2}}~ .
\label{eq:fastness-rad-averaged}
\end{equation}

Tab. 1 shows the time-averaged magnetospheric radius, $\langle
\tilde r_m\rangle$, the ratio $\langle \tilde r_m\rangle/\tilde
r_{\rm cor}$ and the fastness parameter $\langle\omega_s\rangle$
for both models.
 We obtained that
 the time-averaged size of the magnetosphere increases with $\tilde\mu$, and is $\langle r_m
\rangle\approx 5, 5.9$ for models $\tilde\mu30$, and
$\tilde\mu60$, respectively. The fastness parameter also increases
systematically: $\omega_s=3.9, 5.1$ for the same models,
respectively. Both values of $\langle\omega_s\rangle$ are within
the interval of fastnesses, $2.3\lesssim \omega_s \lesssim 16.6$,
expected
in IPs with magnetic fields in the range of $B=10^5-10^6$ G (see
Eq. \ref{eq:fastness-rad-averaged}) \footnote{Note that we
currently cannot model propellers with magnetospheric radii
$\langle r_m\rangle\gtrsim 7 R_s$, because in the cases of large
magnetospheres, the density of the plasma inside the magnetosphere
becomes very low, and the time-step in the simulations becomes too
small. As  a result,  in the models with a corotation radius of
$r_{\rm cor}\approx 2 R_s$, we can only model propellers with
values of $\omega_s\lesssim 6-7$.}.

Simulations performed in dimensionless form provide different
values that can be used for the subsequent development of a
dimensional model of AE Aqr, and for comparisons with
observations. The bottom row of Fig. \ref{2d-fluxes} shows the
matter fluxes onto the star, ${\widetilde{\dot M}}_s$, and into
the wind, ${\widetilde{\dot M}}_w$. One can see that the fluxes
strongly oscillate. Matter accretes onto the star in rare bursts,
because most of time accretion is blocked by the centrifugal
barrier of the rapidly-rotating magnetosphere. Ejections into the
wind (blue line) are more frequent than accretion events (red
line). To characterize each model, we introduce the time-averaged
matter fluxes, which are calculated using a formula similar to Eq.
\ref{eq:timeavg-rm}. The dashed lines show the time-averaged
values of the matter fluxes to the star,
$\langle\widetilde{\dot{M}}_s\rangle$, and to the wind,
$\langle\widetilde{\dot{M}}_w\rangle$.
 Simulations show that most of the inner disc matter is
ejected into the winds.

Tab. \ref{tab:matter-fluxes} also shows the efficiency of the
propeller:
\begin{equation}
f_{\rm eff} = \frac{\langle{{\dot M}_w\rangle}}{\langle {\dot
M}_s\rangle + \langle{\dot M}_w\rangle}. \label{eq:efficiency}
\end{equation}
One can see that propeller efficiency is  $f_{\rm eff}=0.82, 0.85$
for $\tilde\mu=30, 60$, respectively. That is, $82-85$ per cent of
the disc matter is ejected into the wind, while $18\%-15\%$
accretes onto the star \footnote{We should note that our
\textit{axisymmetric} model may provide too high accretion rate
onto a star, because we model the (unknown) rate of penetration of
the disc matter through the magnetosphere by high diffusivity
values in the region $r<7R_s$.}.

Next, we use the data obtained in our dimensionless models to
develop the dimensional models of AE Aqr. To obtain a dimensional
value $A$, we take the dimensionless value $\tilde A$ from the
simulations, and multiply it by the reference value $A_0$ :
$A=\tilde A A_0$.  We derive the reference values $A_0$ using our
standard procedure described in Appendix \ref{app:reference
values}.

\section{Comparisons of model with observations of AE Aqr}
\label{sec:comparisons-observations}

Below, we compare the different values calculated in our model
with the values observed in AE Aqr.

\subsection{Magnetic field of AE Aqr derived from comparisons of modelled and observed spin-down rates}
\label{subsec:magnetic field}

In this section, we compare the spin-down rates obtained in our
numerical model with the observed spin-down rate of $P_{\rm
sd}=5.64\times10^{-14}$ s/s, and derive the possible values of the
magnetic field.

The angular momentum of the star is $L_s=I_s\Omega_s$, where the
value of the moment of inertia of the white dwarf is $I_s=k M_s
R_s^2=1.57\times 10^{50} k_{0.2}
 \rm{g cm^2}$.  The time-averaged spin-down rate can be estimated from
 the relation $\langle\dot L_{\rm sd}\rangle
 = I_s\langle\dot\Omega_{\rm sd}\rangle$, and
 can be re-written in the following form:

\begin{equation}
\langle \dot P_{\rm sd}\rangle=\frac{P_s \langle\dot L_{\rm
sd}\rangle}{I_s\Omega_s}=\frac{P_s^2 \langle\dot L_{\rm
sd}\rangle}{2\pi I_s} \approx \nonumber
\end{equation}
\begin{equation}\approx1.05\times 10^{-15} k_{0.2}^{-1}
{B_5}^2 {\tilde\mu_{60}}^{-2} {\langle\widetilde{\ldot}_{\rm
sd}\rangle}~\frac{s}{s}~. \label{eq:Pdot_sd_dimensional}
\end{equation}
Table \ref{tab:magnetic-fields} shows the spin-down rates
 obtained in our models.

To derive the magnetic field of AE Aqr, we equate the spin-down
rates $\langle\dot P_{\rm sd}\rangle$ obtained in the simulations
(see Eq. \ref{eq:Pdot_sd_dimensional}) with the observed spin-down
rate. We obtain the magnetic field in the following form:
\begin{equation}
B_{\rm Psd}\approx 7.33\times 10^5 {\tilde\mu}_{60} k_{0.2}^{1/2}
{\langle \widetilde{\dot L}_{\rm sd} \rangle}^{-1/2} {\rm G}~.
\end{equation}
Tab. \ref{tab:magnetic-fields} shows the values of the magnetic
field, $B_{\rm Psd}$, which are $B_s\approx 2.8\times 10^5$ and
$B_s\approx 3.3\times 10^5$ G, for models $\tilde\mu30$, and
$\tilde\mu60$, respectively. One can see that the strength of the
magnetic field increases slightly with $\tilde\mu$. It does depend
on the moment of inertia coefficient, $k_{0.2}$, which can vary by
a factor of $\sim$ 2 (depending on the model of the white dwarf).
Overall, these values are in the range of the magnetic field
values estimated for IPs. Note that \citep{ChoiYi2000} estimated a
magnetic field of the white dwarf of $3\times10^5$ G on basis of
the quiescent X-ray and UV emission. This value is very close to
values obtained in our models. Note that this field is much lower
than the field  $5\times 10^7$ G used in other type of models
(e.g., \citealt{IkhsanovEtAl2004}, and references therein).

The time-averaged spin-down power
 of the star:
\begin{equation}
\langle\dot E_{\rm sd}\rangle=\langle \dot L_{\rm sd} \rangle
\Omega_s =\dot L_0\langle\widetilde{\dot L}_{\rm sd}\rangle
2\pi/P_s \approx \nonumber
\end{equation}
\begin{equation}
\approx 1.80\times 10^{32}{\tilde\mu_{60}}^{-2}{B_5}^2
\langle\widetilde{\dot L}_{\rm sd}\rangle ~{\rm erg/s} ~.
\label{eq:E_sd dimensional}
\end{equation}
corresponds to the spin-down rate ${\dot E}_{\rm sd}\approx
6\times 10^{33} I_{50}$ erg/s $\approx 9.4\times 10^{33}$ erg/s,
derived from observations (see Tab. \ref{tab:magnetic-fields}).

These comparisons show that the propeller model offers a good
explanation for the observed spin-down properties of the star.
However, we cannot yet select one specific model over another,
because both models can explain the observations, although at
slightly different values of the stellar magnetic field.

\begin{table*}
\centering
\begin{tabular}{l|ll|ll}
Model                          & $\langle\dot P_{\rm sd}\rangle$ (s/s) & $B_{\rm Psd}$ (G)  & $\langle\widetilde{\dot{L}}_{\rm sd}\rangle$   & $\langle\dot E_{\rm sd}\rangle$ (\rm erg/s)\\
\hline
$\tilde\mu30$                  & 7.06E-15$B^2_5/k_{0.2}$                      &  2.83E5$\sqrt{k_{0.2}}$     & 1.68   & 9.69E33 $k_{0.2}$     \\
\hline $\tilde\mu60$           & 5.20E-15$B^2_5/k_{0.2}$                      &   3.29E5$\sqrt{k_{0.2}}$    & 4.95   & 9.64E33 $k_{0.2}$     \\
\hline
\end{tabular}
\caption{Values of the magnetic field, $B_{\rm Psd}$, derived from
the comparisons of the spin-down rates, $\dot P_{\rm sd}$.
$\langle\dot{E}_{\rm sd}\rangle$ is the calculated value of
spin-down power.}
 \label{tab:magnetic-fields}
\end{table*}

\subsection{Matter fluxes to the star and to the wind}
\label{subsec:matter-fluxes}

We can now calculate the dimensional values of the matter fluxes
onto the star and into the wind. We take the reference value,
$\dot M_0$, from Eq. \ref{eq:matter-flux-refval} and take the
dimensionless values from Tab. \ref{tab:dimensionless values}
to obtain the time-averaged matter fluxes to the star (subscript
`s') and to the wind (subscript `w'):
\begin{equation}
\langle\dot M_{s,w}\rangle=\dot M_0\langle\widetilde{\dot
M}_{s,w}\rangle \approx 3.47\times
10^{15}{\tilde\mu_{60}}^{-2}{B_5}^2 \langle\widetilde{\dot
M}_{s,w}\rangle g/s ~. \label{eq:M_sw_dimensional}
\end{equation}
Substituting in the values of the magnetic field from Tab.
\ref{tab:magnetic-fields}, we obtain the values of the matter
fluxes (see Tab. \ref{tab:matter-fluxes}). The table also shows
the time-averaged total matter flux through the disc: $
\langle\dot{M}_d\rangle = \langle\dot{M}_s\rangle
+\langle\dot{M}_w\rangle$. One can see that the matter flux in the
disc decreases when parameter $\tilde\mu$ increases.

The  matter flux in the disc obtained in our models,
$4.45\times10^{16} \rm{g/s}
\lesssim\langle\dot{M}_d\rangle\lesssim 1.0\times 10^{17}
\rm{g/s}$, is comparable to the mass transfer rates expected in
nova-like CVs with orbital periods of $\sim 10$ hours: $\sim
1.0\times 10^{17}$ g/s (e.g., \citealt{Dhillon1996}).

We use Eq. \ref{eq:Edot_X from M_x} and our values of
$\langle\dot{M}_s\rangle$ to estimate the luminosity associated
with accretion onto the star, $\langle\dot{E}_{\rm s}\rangle$.
Tab. \ref{tab:matter-fluxes} shows the values of luminosity. The
luminosity $\langle\dot{E}_s\rangle$ is more than 10 times larger
than the accretion luminosity deduced from the observations: $\dot
E_{\rm acc}\lesssim 10^{32}$ erg/s. The last column of Tab.
\ref{tab:matter-fluxes} shows the ratio between the accretion
luminosity obtained in our models
and the observed luminosity of $\dot{E}_s\approx
10^{32} \rm{ergs/s}$. One can see that the accretion luminosity
obtained in our models is $6-14$ times higher than the observed
one.

The relatively high accretion rate onto the star may be connected
with the axisymmetry of our model and our high diffusivity value
($\alpha_d=1$) taken at the disc-magnetosphere boundary. At lower
values of $\alpha_d$, the accretion rate is expected to be lower.
We also should note that to support MHD simulations, a small
amount of matter (floor density matter) is added to the parts of
the magnetosphere with the largest values of the
magnetic-to-matter pressure density. This matter may also
contribute to the accretion rate onto a star.

\begin{table*}
\centering
\begin{tabular}{l|llll|ll}
Model           &   $\langle\dot{M}_s\rangle$ ($\rm g/s$) & $\langle\dot{M}_w\rangle$ (g/s)  & $\langle\dot{M}_d\rangle$ (g/s)  & $f_{\rm eff}$ & $\langle\dot{E}_s\rangle$ (\rm erg/s)   & $\langle\dot{M}_s\rangle/\dot{M}_{\rm obs}$ \\
\hline
$\tilde\mu30$  &   1.89E16$k_{0.2}$                & 8.41E16$k_{0.2}$                   & 1.03E17$k_{0.2}$              & 0.82          & 2.88E33$k_{0.2}\eta$                        & $14.4 k_{0.2}\eta_{0.5}$         \\
\hline
$\tilde\mu60$  &   8.24E15$k_{0.2}$                & 4.65E16$k_{0.2}$                   & 5.47E16$k_{0.2}$              &0.85           & 1.25E33$k_{0.2} \eta$                       & $6.24 k_{0.2}\eta_{0.5}$        \\
\hline
\end{tabular}
\caption{The time-averaged matter fluxes onto the star,
$\langle\dot{M}_s\rangle$,  into the wind,
$\langle\dot{M}_w\rangle$, and in the disc,
$\langle\dot{M}_d\rangle$. $f_{\rm eff}$ is the efficiency of the
propeller.
 $\langle\dot{E}_s\rangle$
is the energy flux associated with accretion onto the star and
calculated using Eq. \ref{eq:Edot_X from M_x}; ${\dot M}_{\rm
obs}$ is calculated using the observed luminosity: $\dot{E}_{\rm
obs}\approx 10^{32}$ erg/s.} \label{tab:matter-fluxes}
\end{table*}

\begin{figure*}
\centering
\includegraphics[width=16.0cm]{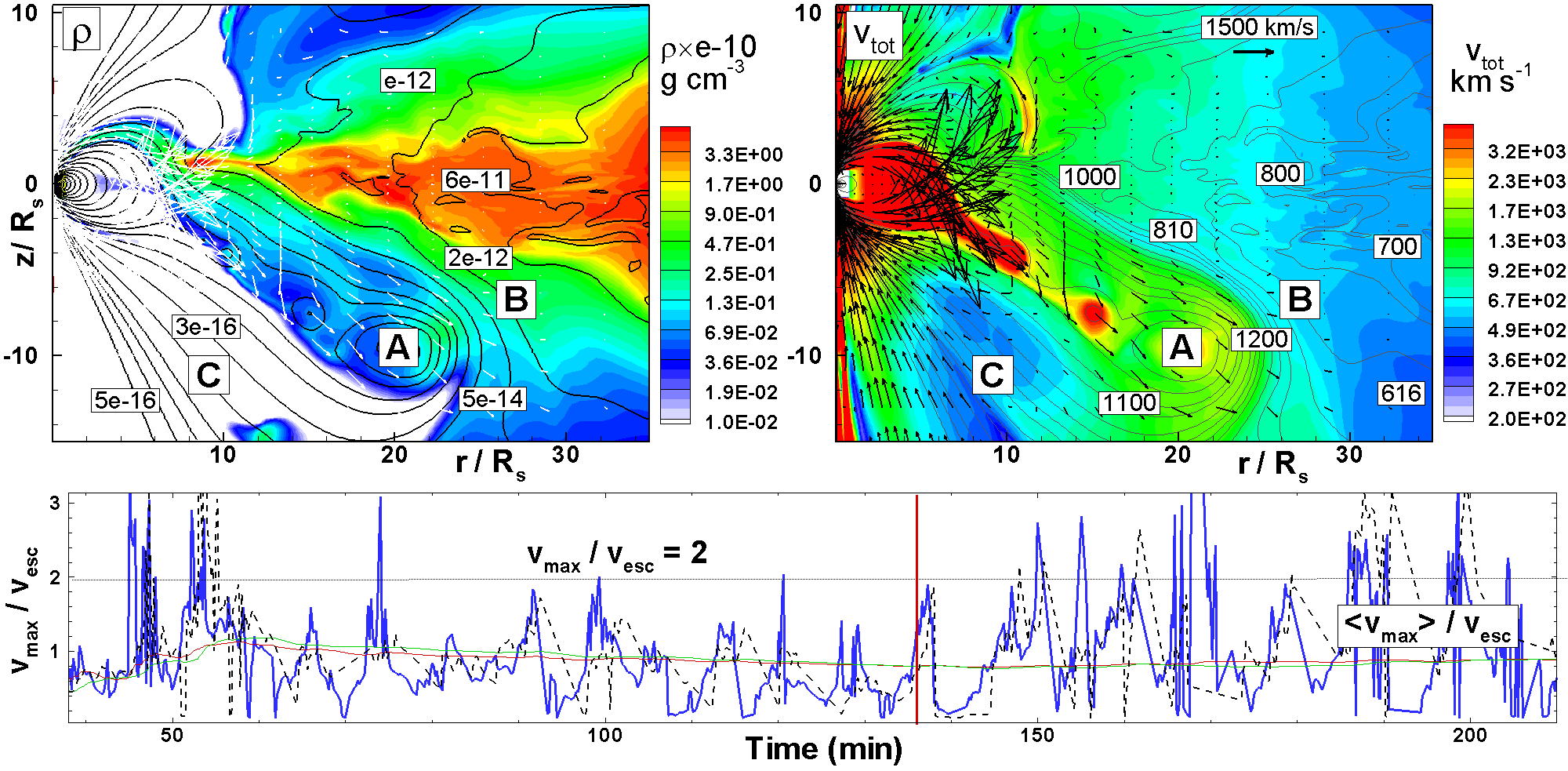}
 \caption{\textit{Top panels:} Different characteristics of the flow during propeller
 ejection (at $t=188$ min). Black lines are sample poloidal field lines, arrows are proportional to velocity
 vectors. The background in the left and right panels shows the density, $\rho$, and the absolute value of the total velocity, $v_{\rm tot}$, respectively.
\textit{Bottom panel:}
 Temporal variation
 of the normalized maximum poloidal velocity, taken at cylinders with radii $r=20 R_s$ (blue line) and $r=30 R_s$ (dashed black line).
 Red and green lines  show the time-averaged values, respectively. Vertical line shows the moment of time corresponding to top panels.} \label{2d-rho-vel-3}
\end{figure*}

\subsection{Velocities of matter in the wind}
\label{subsec:velocities}

\citet{WelshEtAl1998} performed high-speed spectrophotometric
observations of AE Aqr in the $H_\alpha$ spectral line and found
that the spectral line  varies rapidly in both red and blue wings.
Doppler signatures from variable lines show that the red and blue
wings span from $250$ km/s to $1,600$ km/s, with a significant
amount of radiation which comes from the matter that flows at
radial velocities of $\lesssim 500$ km/s, and a smaller amount of
matter flowing at higher velocities (see Figures 8-10 from
\citealt{WelshEtAl1998}). Below, we analyze the distribution of
velocities in our model, and discuss the possible locations where
this flaring radiation may originate.

\begin{figure*}
\centering
\includegraphics[width=18.0cm]{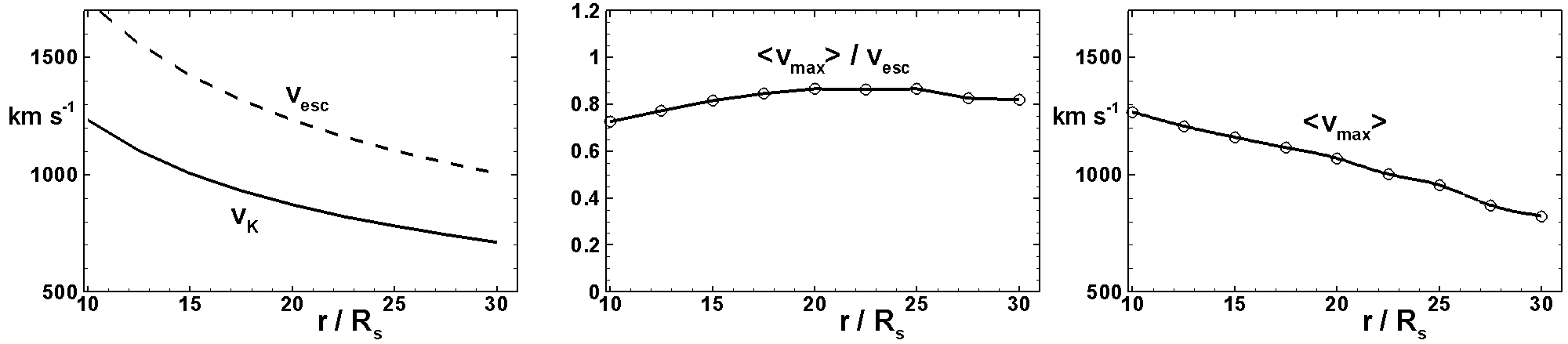}
 \caption{\textbf{\textit{Left panel:} Variation of Keplerian, $v_K$, and
 escape velocity, $v_{\rm esc}$, with the distance from the star.
 \textit{Middle panel:} Variation of the time-averaged maximum poloidal velocity normalized by the local escape velocity with
 the distance from the star. \textit{Right panel:} Variation of the time-averaged maximum poloidal velocity with the distance from the star.}}
 \label{vmax-rad-3}
\end{figure*}

First, we calculate the total velocity of matter flow in different
parts of our simulation region, $v_{\rm
tot}=\sqrt{v_r^2+v_z^2+v_\phi^2}$, where  $v_r$, $v_z$ and
$v_\phi$ are the components of velocity in $r,z$ and $\phi$
directions, respectively. Fig. \ref{2d-rho-vel-3} (top right
panel) shows the distribution of total velocity during a typical
propeller ejection.  One can see that the total velocity is high
in the area of propeller ejecta, $v_{\rm tot}\sim 1,000-1,200
\rm{km/s}$ (see region A in the plot), and is lower in region B,
between the disc and the ejecta.  Velocity is lowest in the disc,
with a minimum value of  $v_{\rm tot}\approx (650-700) \rm{km/s}$.
These velocities are too high to explain the low-velocity
component observed in the Doppler shifts of the $H_\alpha$
spectral line. The lowest velocity is associated with Keplerian
velocity of the disc matter. This velocity decreases
 with distance from the star as
$v_K=\sqrt{GM_s/R_s}\approx 872 (r/20)^{-0.5}$ (see also left
panel of Fig. \ref{vmax-rad-3}), and reaches the value of  $v_\phi
\sim 500 \rm{km/s}$ at the distance of $r\approx 61 R_s$,  which
is approximately twice as large as our simulation region in the
radial direction.

Next, we investigate the poloidal velocities,
$v_p=\sqrt{v_r^2+v_z^2}$. Fig.  \ref{2d-rho-vel-3} shows that the
poloidal velocities (vectors in the plot) are largest in the area
of the propeller ejecta (region A in the plot). They are smaller
in the disc wind (region B). There is also region C (closer to the
axis), where velocities may be very high. However, the density in
this region is very low (see distribution of densities in the top
left panel of the same figure), and this region does not
contribute much to the matter flux. Analysis of the poloidal
matter fluxes, $\rho v_p$, shows that most of the matter flows in
the regions  A (higher velocities) and B (lower velocities).

The question arises whether the ejected matter moves with
super-escape velocities.  For this analysis, we study the
distribution of poloidal velocities with distance from the star.
We surround a star with cylindrical boxes of radii $r$ and heights
$z=\pm r$, and search for the maximum poloidal velocity, $v_{\rm
max}$, at the cylindrical boxes $(r,z=\pm r)$ \footnote{We
de-selected regions of very high velocity with low matter flux, by
placing condition that the matter flux should be larger some small
value,
 $\rho v_p=0.001$ (in dimensionless
units) (see details in \citet{RomanovaEtAl2018} and Fig. 7 from
this paper, which shows the position of the maximum poloidal
velocity.}. The bottom panel of Fig. \ref{2d-rho-vel-3} shows
variation of the maximum poloidal velocity with time at the
cylindrical boxes with radii $(r,z)=(20R_s,\pm20R_s)$ (solid blue
line) and $(r,z)=(30R_s,\pm30R_s)$ (long-dashed line). One can see
that the maximum velocity varies between  small values, much
smaller than the escape velocity, and large values, which are 2-3
times larger than the escape velocity. This analysis shows that
the ejected matter is not gravitationally bound, and will continue
to move further away to larger distances from the star.

Another question is whether matter moves from the star with
acceleration or deceleration. To analyze the variation of poloidal
velocity with distance from the star, we  calculate the
time-averaged maximum poloidal velocity, $\langle v_{\rm
max}\rangle$ (see details in \citealt{RomanovaEtAl2018}). The
bottom panel of Fig. \ref{2d-rho-vel-3} shows that the normalized
time-averaged velocities, $\langle v_{\rm max}\rangle/v_{\rm
esc}$, are approximately the same in the cases of smaller and
larger boxes. We also calculated $v_{\rm max}/v_{\rm esc}$ and
$\langle v_{\rm max}\rangle/v_{\rm esc}$ at different cylindrical
boxes and observed that the ratio $\langle v_{\rm
max}\rangle/v_{\rm esc}$ stays approximately constant (see middle
panel of Fig. \ref{vmax-rad-3}). The escape velocity decreases
with distance, and therefore the poloidal velocity also decreases
with distance. The  right panel of Fig. \ref{vmax-rad-3} shows
that the poloidal velocity decreases with distance approximately
linearly. Approximating this dependence to larger distances from
the star, we obtain  $\langle v_{\rm max}\rangle\approx  500 $
km/s at the distance of $r\approx 60 R_s$. Overall, both the
azimuthal and poloidal velocities decrease with distance and are
expected to have lower values at larger distances from the star.

Propeller ejections are non-stationary and shock waves may form,
where new ejecta interacts with a slower-moving matter of the
earlier ejected matter. We suggest that the flares
 observed in optical, UV and X-ray spectral
bands may be associated with radiation in these shocks and in the
surrounding medium.   To explain the relatively low velocities in
radiating matter, one should consider distances that are two times
(or more) greater than  compared with our simulation region.

\begin{figure}
\centering
\includegraphics[width=8.0cm]{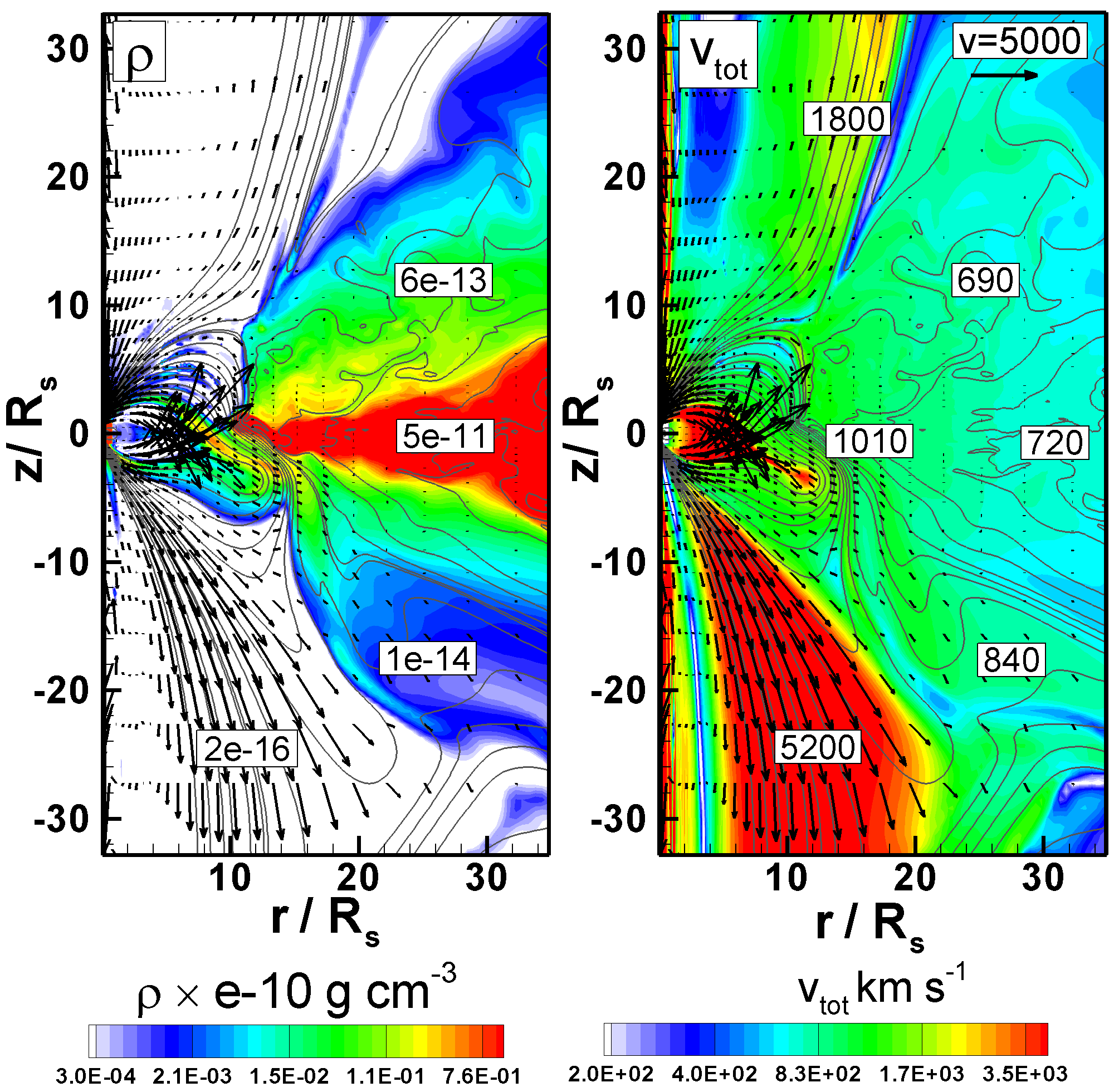}
 \caption{An example of outflows into the magnetic, Poynting flux jet. The meaning of the color background, vectors and lines
 is the same, as in top panels of Fig. \ref{2d-rho-vel-3}.} \label{2d-poynt-2}
\end{figure}

We should note that  a small amount of matter is rapidly
accelerated by the magnetic pressure force in regions, closer to
the axis (see Fig. \ref{2d-poynt-2}). These magnetic (Poynting
flux) ejections do not correlate directly with the matter
ejections. They occur during episodes of strong inflation of the
magnetic loops.
 Ejections are non-stationary, and formation of shock waves
is also expected. In shocks, the magnetic energy can be converted
into the particle energy due to reconnection or other processes
(e.g., \citealt{RomanovaLovelace1992},
\citealt{RomanovaLovelace1997}).
 These shocks may be responsible
for radio flares and for the high-energy radiation.  These
magnetic outflows are magnetically-collimated and are expected to
form the jet-like flow at larger distances from the star. They are
similar to magnetic jets studied in compact stars and quasars
(e.g., \citealt{UstyugovaEtAl2000,TchekhovskoyEtAl2012}). We
suggest that flaring radiation in radio band may originate in the
shock waves of this jet. The similarity between the radio spectra
of AE Aqr and that of microquasar Cyg X-3  is in favour of this
hypothesis. Some part of the magnetic energy can be converted to
the high-energy radiation. However, it can be a small, and
therefore there is no contradiction with recent luck in finding
TeV radiation from AE Aqr (e.g., \citealt{AlexicEtAl2014}).

\begin{table}
\centering
\begin{tabular}{l|lll}
 Model             & $\langle\widetilde{\dot{E}_f}\rangle$ & $\langle\widetilde{\dot{E}_m}\rangle$  & $\langle\dot{E}_f\rangle$ (${\rm erg/s}$) \\
\hline
$\tilde\mu30$      &  0.59          &        0.07         &  1.0E34$k_{0.2}$                  \\
\hline
$\tilde\mu60$      &  1.59          &  0.09&  9.1E33$k_{0.2}$                                         \\
\hline
\end{tabular}
\caption{Time-averaged energy fluxes to the wind carried from the
surface of the star by the magnetic field and my matter.
$\langle\widetilde{\dot{E}_f}\rangle$ and $\langle\dot E_f\rangle$
are dimensionless and dimensional values, respectively.}
\label{tab:energy fluxes}
\end{table}

\subsection{Time intervals between flares}
\label{subsec:time intervals}

Next, we compare the variability time scales obtained in our
numerical models with the time scales observed in the light curves
of AE Aqr. Fig. 3 from \citet{MaucheEtAl2012} shows that there are
two types of flares: some flares occur rarely, with a time
interval of a 1-3 hours between each flare (see individual large
flares marked by vertical dashed lines); other flares occur more
frequently, on time intervals of 10-30 minutes (see multiple
flares on the left-hand side of Fig. 3 from
\citealt{MaucheEtAl2012}, and also the sub-flares that occur on
other time intervals). We can compare the observed time scales
with the time scales obtained in our models.

Fig. \ref{2d-fluxes} shows the variability obtained in model
$\tilde{\mu}60$. The bottom
 panel shows the matter fluxes onto the star, $\dot M_s$, and
into the wind, $\dot M_w$. The blue line in the top panel shows
that the strongest ejections into the wind occur every 10-20
minutes. This time scale is similar to the time scales of the
frequent flares in the observed light curve. Ejections into the
wind are slightly more frequent ($\sim 5-10$ min) in the model
with the smaller magnetosphere, $\tilde\mu30$. Overall, the
variability time scales match the short time scale oscillations in
the observed light curve.

In our simulations, ejections into the wind occur in brief bursts,
because outflows are only possible if the field lines connecting
the star with the inner disc inflate and open. Inflation becomes
possible when a sufficient amount of matter accumulates at the
inner disc and diffuses through the field lines of the outer
magnetosphere.
 Most of the time, matter is accumulated at the inner disc, while the
outflow events are relatively brief. This is why the variability
curve looks spiky (see bottom panel of Fig. \ref{2d-fluxes}).
However, the  small scale oscillations observed in AE Aqr are not
spiky.
 We suggest that a proper inclusion of radiation in our
model may change the shape of the variability curve, possibly
making the oscillations less spiky.

On the other hand, the variability curve associated with the
variation of the inner disc radius, $r_m$ (see middle row of Fig.
\ref{2d-fluxes}), is not spiky. It strongly resembles the observed
light curve for short time scale flares.
 In some theoretical models, it is suggested
that the flaring radiation is connected with the processes at the
disc-magnetosphere boundary and in the inner disc, such as the
heating in the turbulent layer of the boundary
\citep{PapittoTorres2015}, or the acceleration of particles in the
magnetized plasma of the inner disc \citep{MeintjesdeJager2000}.
If the radiation really does originate at the inner disc, then the
oscillations of the inner disc (observed in our model) will lead
to the variations in the observed light curves.

The origin of the less frequent flares in the light curves of AE
Aqr may be connected with the non-stationary accretion
 expected in the propeller regime.
On the other hand, non-stationary ejections can be due to the
non-stationary accretion of matter from the secondary star, which
may have the form of inhomogeneous streams, or blobs (e.g.,
\citealt{WynnKing1995}). More detailed analysis of these processes
is beyond the scope of this paper.

\subsection{Energy in Flares}
\label{subsec:energies}

A star loses its rotational energy. A part of this energy is
transferred to the energy of the inner disc winds, while another
part flows from polar regions of the star in the form of magnetic,
Poynting flux jets. The time-averaged energy fluxes carried by
matter (m) and by the magnetic field (f) through some surface
$S(r,z$) are:
\begin{eqnarray} \langle\dot E_{m,f}\rangle\approx
 5.29\times 10^{32}{\tilde\mu_{60}}^{-2}{B_5}^2 \langle\widetilde{\dot E}_{m,f}\rangle~  {\rm erg}/{\rm s}
 ~,
\label{eq:energy-wm}
\end{eqnarray}
where $\langle\widetilde{\dot E}_{m,f}\rangle$ are corresponding
dimensionless values. Table \ref{tab:energy fluxes} shows energy
fluxes calculated at the stellar surface. One can see that almost
all energy is carried by the magnetic field. A smaller amount of
energy is carried by the matter, which accretes onto the star.

Observations of AE Aqr show aperiodic flares in optical, UV, X-ray
and radio bands (e.g., \citealt{Patterson1979}). Similar flares
are observed in the UV spectrum \citep{EracleousHorne1996}. The
luminosity of flares in the high state in the Balmer continuum
reaches the value of $8.4\times10^{31}$ erg/s, and  values a few
times smaller  in the low state in the Balmer continuum, and in
the UV and Blamer spectral lines (see Tab. 3 from
\citealt{EracleousHorne1996}). The X-ray luminosity is lower,
while the radio luminosity is much lower. \footnote{Energetic
flares in $\gamma$-rays  were reported \citet{MeintjesEtAl1994},
however were not confirmed by other groups.}

In our model, matter is ejected from the inner disc into the wind
in non-stationary bursts (see bottom panel of Fig.
\ref{2d-fluxes}). The ejected matter is  accelerated by the
magnetic pressure force from zero velocity (near the disc) up to
hundreds of km/s at larger distances (see Sec.
\ref{subsec:velocities}). Accelerated chunks of matter, wile
interacting with earlier ejected matter, may form shock waves, and
a significant part of the kinetic energy can be radiated in the
shock waves and observed as flares\footnote{In our code, the
energy equation is written in the entropy form, and the shocks
cannot be modelled.}.

Matter is ejected from the inner disc in chunks of different mass.
Using a plot in the bottom panel of Fig. \ref{2d-fluxes}, we can
estimate the mass ejected in individual bursts. For example, at
time $t=123$ min, the large ejection lasted for $\Delta t\approx
120$ sec and the total integrated ejected mass is estimated as
$\Delta M_{\large}\approx 3\times 10^{19}$ g. Smaller ejections
have mass of $\Delta M \sim (0.3-1.0)\times 10^{19}$ g.

The kinetic
energy, carried by individual chunks is
$$
E_{\rm kin}\approx\frac{1}{2}\Delta M v_p^2\approx 3.7\times
10^{34} \frac{\Delta M}{3\times 10^{19} g} \bigg(\frac{v_p}{500
\rm{km/s}}\bigg)^2~ \rm{erg}
$$
A typical duration of strong ejections is $\Delta t\approx 2$ min,
and the rate of energy release during ejection is
$$
\dot{E}_{\rm kin}\approx\frac{E_{\rm kin}}{\Delta t} \approx
3.1\times 10^{32}\frac{\Delta M}{3\times 10^{19} g}
\bigg(\frac{v_p}{500 \rm{km/s}}\bigg)^2\frac{2 \rm{min}}{\Delta
t}~ \frac{\rm erg}{\rm s} ~.
$$
\noindent This rate of energy release  is sufficient to explain
the observed flares.

Our model can be compared with the model of colliding blobs (e.g.,
\citealt{WynnEtAl1997}). In this model, it is suggested that
matter accretes onto the magnetosphere in blobs, which are
subsequently ejected by the propelling magnetosphere in clumps,
which collide with each other and radiate (e.g.,
\citealt{PearsonEtAl2003,ZamanoEtAl2012}). To explain the observed
flares, they estimate the mass involved in collisions as $\Delta
M_{\large}\approx 3\times 10^{19}$ g and the total energy,
released during collisions, as  $E\approx 3\times 10^{33}$ erg/s.
It is interesting, that the mass of our largest ejections is
similar to the mass in their model. However, they considered
longer-lasting flares, while our flares are relatively brief.

In our model, energy also flows in the form of invisible magnetic
energy. Some energy is associated with the magnetic field lines
expanding together with matter into the inner disc wind. On the
other hand, a significant amount of magnetic energy  flows from
the surface of the star along the inflated field lines
 in the form of a Poynting flux jet, where a small
amount of matter is accelerated rapidly by the magnetic force. The
Poynting flux
 outflows are non-stationary, and particles may
be accelerated at the magnetic shocks (e.g.,
\citealt{RomanovaLovelace1997}), possibly up to very high
energies. However, it is not clear which part of the energy will
propagate in the form of an invisible, magnetic jet and which part
will be converted to accelerated particles and radiation
\footnote{Recent observations of the very high-energy radiation
from AE Aqr, performed by \textit{MAGIC} at the energies of $>
100$ GeV \citep{AlexicEtAl2014} and by \textit{Fermi}-LAT in the
100 MeV-300 GeV energy range, have shown that this radiation, if
present, should be weak, much weaker than that of $\sim 10^{34}$
erg/s reported in the earlier observations in the TeV band (e.g.,
\citealt{deJagerEtAl1986}). In our model, individual ejections
into the Poynting flux jet may have the sufficiently high energy
flux of $\sim  10^{34}$ erg/s. However, only a small part of this
energy can be converted into the very high-energy radiation.}.

\section{Summary}
\label{sec:conclusions}

In this work, we have developed a propeller model of AE Aqr using
axisymmetric simulations. We suggested that some type of an
accretion disc forms around the white dwarf and interacts with the
magnetosphere of the star in the propeller regime. We compared the
results of our models, $\tilde\mu30$ and $\tilde\mu60$, with the
observations. In these models, the time-averaged magnetospheric
radii are $\langle r_m\rangle\approx 5$ and $6$, and the fastness
parameter values are $\omega_s=3.9$ and $5.1$, respectively. Our
conclusions are the following:

\smallskip

\textbf{1.} Both models can explain the rapid spin-down of AE Aqr,
although at slightly-different values of the magnetic field of the
white dwarf: $B_s\approx (2.9, 3.3)\times10^5$ G for
$\tilde\mu=30$ and $60$, respectively.

\smallskip

\textbf{2.} In both models, the disc-magnetosphere interaction is
a strongly non-stationary process, where the inner disc
oscillates. The total time-averaged accretion rate in the disc
 is $\dot M_d=(1.0, 0.55)\times 10^{17}$ g/s in the above
 two models, respectively.

\smallskip

\textbf{3.} Most of the inner disc matter is ejected into
conically-shaped winds, and a much smaller part accretes onto the
star. Our axisymmetric model is not precise in providing the
accretion rate onto the star, due to unknown diffusivity rate.

\smallskip

\textbf{4.} The main flaring variability can be explained through
the processes in the non-stationary outflows (possibly by the
radiation in shocks). The time-averaged total energy budget in the
outflows, $2\times10^{33}$ erg/s, is sufficient for explaining the
flares observed in different wavebands.

\smallskip

\textbf{5.} The predicted  accretion rate onto a star is higher
than that obtained from the observations of accretion luminosity
of AE Aqr ($E_{\rm acc}\lesssim 10^{32}$ erg/s). The relatively
high accretion rate in our models can be explained by the high
diffusivity at the disc-magnetosphere boundary, taken in our
axisymmetric models.

\smallskip

\textbf{6.} Velocities of matter in the simulation region
 $v\sim 600-1,200$ km/s, are high and cannot explain the low velocity component
 of the flaring radiation observed in  the $H_\alpha$ spectral
line. Matter responsible for this flaring radiation should be
located  at least at twice as large distance from the star,
compared with our simulation region.

\smallskip

\textbf{7.} Both accretion and ejections are non-stationary and
occur in brief episodes. The ejections of matter into the outflows
and the oscillations of the inner disc occur on a time scale of
$\Delta t\approx 10-20$ minutes. This variability matches the
short time scale variability observed in the light curves of AE
Aqr. The longer time scale variability may be connected with the
non-stationary accretion from the disc.

\smallskip

Overall, the developed models of AE Aqr are in reasonable
agreement with the observational data. One of the
 inconsistencies is the presence of an
accretion disc in our model and no observational evidence of an
accretion disc in the Doppler tomograms (e.g.,
\citealt{WynnEtAl1997}). We should note that, in our models of the
strong propeller regime, the inner disc strongly oscillates. This
may lead to a variable, non-ordered disc, which can be difficult
to detect using the Doppler tomography technique (which suggested
a steady flow of matter in the equatorial plane of the binary,
e.g., \citealt{Echevarria2012}).  Our propeller models work even
in the cases where the disc is variable, relatively small, or when
it forms as a temporary feature. Additionally, in our model, most
of the disc  matter is ejected into the outflows, which can
distort information about the accretion disc in the Doppler
tomograms.

Another important inconsistency of the model is the fact that the
model provides a higher accretion rate onto the surface of a star
compared with that derived from the observations. We should point
out that our axisymmetric model we suggested a high diffusivity in
the inner part of the simulation region (to mimic the 3D
instabilities), which may lead to a higher accretion rate. On the
other hand, in more realistic 3D simulations, the magnetic axis
can be tilted about the rotational axis, which may lead to lower
accretion rate onto the surface of the star in the propeller
regime.

\section*{Acknowledgments}
The authors thank the anonymous referee for valuable comments and
corrections, and Gagik Tovmassian for helpful discussion.

Resources supporting this work were provided by the NASA High-End
Computing (HEC) Program through the NASA Advanced Supercomputing
(NAS) Division at the NASA Ames Research Center and the NASA
Center for Computational Sciences (NCCS) at Goddard Space Flight
Center. The research was supported by NASA grant NNX14AP30G. AVK
was supported by the RFBR grant 18-02-00907.

{}

\appendix

\section{Description of Numerical Model}
 \label{app:numerical-model}

\subsection{Initial and boundary conditions}
\label{app:initial and boundary}

 The initial conditions are similar
 to those
 used in our previous work \citep{RomanovaEtAl2018}, where
 the initial density and entropy distributions
 were calculated
 by balancing the gravitational, centrifugal and
pressure forces. The disc is
 initially cold and dense, with temperature $T_d$ and density $\rho_d$. The corona
 is hot and rarified, with temperature $T_c = 3\times 10^3 T_d$ and density $\rho_c = 3.3\times 10^{-4} \rho_d$.
 In the beginning of
 the simulations,
 the inner edge of the disc is placed
 at $r_d$ = 10, and the star rotates with $\Omega_i$ = 0.032 (corresponding to $r_{\rm
cor}$ = 10), so that the
  magnetosphere and the inner disc initially corotate. This condition helps to ensure that
  the magnetosphere and the disc are initially in near-equilibrium at the disc-magnetosphere
  boundary.
  The star is gradually spun up from $\Omega_i$ to the final
  state with angular velocity $\Omega_s\approx 0.35$, which corresponds to $r_{\rm cor}=2$.

Initially, the disc is threaded by the dipole magnetic field of
the star. We also add a small ``tapered'' poloidal field inside
the disc. This tapered field helps initialize the MRI in the disc
and has the same polarity as the stellar field at the
disc-magnetosphere boundary. To initialize the MRI, 5\% velocity
perturbations are added to $v_\phi$ inside the disc.

he boundary conditions are identical to those described by
\citet{LiiEtAl2014}.

\subsection{Grid and code description}
\label{app:grid and code}

 We use a Godunov-type code to solve the MHD equations in cylindrical
coordinates \citep{KoldobaEtAl2016}. The axisymmetric grid is in
cylindrical ($r$, $z$) coordinates with mesh compression towards
the equatorial plane and the $z$-axis, so that there is a larger
number of cells in the disc plane and near the star. In the models
presented here, we use a non-uniform grid with $190 \times 306$
grid cells, corresponding to a grid that is 36 by 66 stellar radii
in size.

\subsection{Reference values}
\label{app:reference values}

 We find the reference pressure
from
the relationship $p_0 = B_0^2$, where $B_0=B_s/\tilde{\mu}$ is the
reference magnetic field ($B_s$ is the magnetic field of the star
and $\tilde\mu$ is the dimensionless magnetic moment of the star).
The reference density is $\rho_0=p_0/v_0^2$, and the reference
temperature is $T_0 = {\cal R} p_0/\rho_0$, where $\cal R$ is the
Rydberg constant. Table
 \ref{tab_ref} shows different reference values and their dependencies on $B_s$ and $\tilde\mu$.

Using the main reference values, we can calculate the reference
values for the fluxes: matter flux, $\dot M_0=\rho_0 v_0 R_0^2$,
angular momentum flux, $\dot L_0 = \dot M_0 v_0 R_0$, and energy
flux, $\dot E_0 = \mdot_0 v_0^2$. Taking into account the fact
that $\rho_0 v_0 R_0^2=B_0^2/v_0^2=({B_s}/{\tilde\mu})^2
(R_0^2/v_0)$, we obtain the reference fluxes in the following
form:

\begin{eqnarray} \dot M_0
 \approx 3.47\times 10^{15}{\tilde\mu_{60}}^{-2}{B_5}^2
 g/s ~, \label{eq:matter-flux-refval} \end{eqnarray}
\begin{eqnarray} \dot L_0 \approx
 9.53\times 10^{32}{\tilde\mu_{60}}^{-2}{B_5}^2 ~ {\rm erg} ~,
\label{eq:angmom-flux-refval}~\end{eqnarray}
\begin{eqnarray} \dot E_0\approx
 5.29\times 10^{32}{\tilde\mu_{60}}^{-2}{B_5}^2 ~  {\rm erg}/{\rm s}
 ~.
\label{eq:energy-flux-refval}
\end{eqnarray}

\begin{table}
\centering
\begin{tabular}{lll}
\hline {\bf Parameters} &     Reference Values         \\
\hline
$M_0$ [\msun]         & 0.8                          \\
$R_0$ [cm]         & $7\times10^{8}$                  \\
$v_0$ [km s$^{-1}$]& $3.90 \times 10^{3}$             \\
$P_0$ [s]          & 11.26                            \\
$B_0$ [G]          & $1.67 \times 10^3 {\tilde\mu_{60}}^{-1} B_5 $         \\
$\rho_0$ [g cm$^{-3}$] & $1.82 \times 10^{-11} {\tilde\mu_{60}}^{-2} B_5^2 $    \\
$n_0$ [1 cm$^{-3}$] & $1.09 \times 10^{13}     {\tilde\mu_{60}}^{-2} B_5^2 $         \\
$T_0$ [K]                  & $1.83 \times 10^9$                                              \\
$T_{\rm disc,0}$ [K] & $6.11 \times 10^5$                                 \\
\hline
\end{tabular}
\caption{Reference values for CV AE Aqr.  Many of the reference
values depend on the (unknown) value of the magnetic field of the
star, $B_s$, and the dimensionless magnetic moment
$\tilde{\mu}=60$. } \label{tab_ref}
\end{table}

\end{document}